\documentclass[twocolumn,prb]{revtex4}
\usepackage{epsfig,amssymb,amsmath}
\begin{document}
\title{Investigation of the breakpoint region in stacks with a finite number of intrinsic Josephson junctions}

\author{ Yu. M. Shukrinov~$^{1}$}
\author{F. Mahfouzi~$^{2}$ }
\author{N. F. Pedersen~$^{3}$ }

\address{$^{1}$ BLTP, JINR, Dubna, Moscow Region, 141980, Russia \\
$^{2}$Institute for Advanced Studies in Basic Sciences, P.O.Box 45195-1159, Zanjan,
Iran\\
$^3$Oersted-DTU, Section of Electric Power Engineering, Technical University of Denmark,
DK-2800, Lyngby, Denmark }

\date{\today}

\begin{abstract}
We study the breakpoint region on the outermost branch of current-voltage characteristics of the stacks with
different number of intrinsic Josephson junctions. We show that at periodic boundary conditions the breakpoint
region  is absent for stacks with even number of junctions. For stacks with odd number of junctions and for
stacks with nonperiodic boundary conditions the breakpoint current is increased with number of junctions and
saturated at the value corresponding to the  periodic boundary conditions. The region of saturation and the
saturated value depend on the coupling between junctions. We explain the results by the parametric resonance at
the breakpoint and excitation of the longitudinal plasma wave by Josephson oscillations. A way for the
diagnostics of the junctions in the stack is proposed.
\end{abstract}
\maketitle

\section{Introduction}
A series of experiments devoted  to intrinsic Josephson junctions\cite{muller1} (IJJ) shows the growing interest
in the current-voltage characteristics (IVC) of the finite stack.\cite{bae,yurgens,doh} Different kinds of
couplings between intrinsic Josephson junctions determine a variety of the IVC observed in HTSC and different
models are exploited for their descriptions. Among them are the inductive \cite{ped93} and capacitive coupling
\cite{koyama96,machida99} models, which usually give approximately similar results. A unified theory for
magnetic and electric coupling in multistacked Josephson junctions was developed.\cite{machida04} The
capacitively coupled Josephson junction model (CCJJ) seizes the main dynamical properties of the IJJ system,
describes the multibranch structure in the IVC of the stack of IJJ\cite{koyama96,sm-physC1} and explains the
microwave resonant absorption.\cite{matsuda} On the other side a diffusion current plays an important role in
the stack of IJJ \cite{ryndyk}. The CCJJ model with diffusion current (CCJJ+DC model) was derived in
Ref.\onlinecite{machida00} on the microscopic level. It gives an equidistant branch structure.\cite{sm-physC2}
Close to the hysteresis jump the system of IJJ is unstable towards a switching. A resonance between the
Josephson and plasma oscillations causes the system to switch  to another branch. This mechanism in case of one
Josephson junction has been considered long time ago.\cite{ped78}

The  "breakpoint region" (BPR) on the IVC of the stack of IJJ was demonstrated in Ref.\onlinecite{{sm-sust1}}
and it is explained as a result of resonance between Josephson and plasma oscillations. We consider, that
simulation of the IVC of IJJ by different groups using different models shows  the BPR on the outermost branch
as well, but the authors did not mention it (see particularly,  Fig.3a in Ref.\onlinecite{machida99} in CCJJ
model; Fig.1 in Ref.\onlinecite{ryndyk} in charge imbalance (CIB) model; Fig.2(left) in Ref.\onlinecite{sm-jpcs}
in CIB model). To our knowledge, no precise experiment to observe the BPR has been carried out yet.

In this paper we show that a detailed investigation of the breakpoint current $I_{bp}$ and BPR width $w_{bp}$
gives us  new important information concerning the creation of longitudinal plasma waves (LPW) in stacks of IJJ
and the peculiarities of the stacks with a finite number of IJJ. We study the IVC of IJJ  in the framework of
the CCJJ+DC model and investigate the dependence of the $I_{bp}$ and the $w_{bp}$ on the number of junctions N
in the stack. We demonstrate the existence of the BPR on the branches corresponding to the stack with one
oscillating junction (O-junction) and show that such information may allow to develop a new method for the
diagnostics of the IJJ.

\section{Model and method}

In the CCJJ model\cite{koyama96} a relation between the charge $\rho_l$ and the generalized scalar potential
$\Phi_l$ of the $l$-th layer is $\rho_l=-\frac{1}{4\pi r^{2}_D}\Phi_l$, where $r_D$ is Debye screening length
and $\Phi_l$ is expressed through a scalar potential $\phi_l$ and derivative of phase $\theta_l$ of
superconducting condensate by $\Phi_l(t)=\phi_l - \frac{\hbar}{2e}\frac{\partial \theta_l}{\partial
t}$.\cite{koyama96, ryndyk} The last relation reflects a nonequilibrium nature of the ac Josephson effect in
layered superconductors\cite{ryndyk}. The superconducting layers are in nonequilibrium state due to the
injection of quasiparticles and Cooper pairs. In the equilibrium state $\Phi_l(t)=0$. In the CCJJ+DC
model\cite{sm-physC2} with diffusion current $J_D^l=-\frac{\Phi_{l}-\Phi_{l+1}}{R}$  between layers $l$ and
$l+1$ the total external current through the stack has a form
\begin{eqnarray}
J=C\frac{dV_{l}}{dt}+J_c^l\sin(\varphi_{l})+\frac{\hbar}{2eR}\dot{\varphi}_{l}, \label{current}
\end{eqnarray}

where $V_l$ is the voltage between superconducting layers $l+1$ and $l$ (see below),  $\varphi_{l}$ is the
gauge-invariant phase difference $\varphi_l(t)= \theta_{l+1}(t)-\theta_{l}(t)-\frac{2e}{\hbar}\int^{l+1}_{l}dz
A_{z}(z,t)$ between layers $l+1$ and $l$, $R$ is junction's resistance, $A_z$ is the vector potential in the
barrier. This total external current is different from the current in the CCJJ model by third term in the right
hand side of the equation ~(\ref{current}). In the CCJJ model it is equal to $V_l/R$. As a result, in the
CCJJ+DC model we obtain the following system of dynamical equations for the phase differences $\varphi_{l}$

\begin{eqnarray}
\partial^2\varphi_l/\partial t^2 =\sum\sb{l'}\,A_{ll'}[I-\sin\varphi_{l'}-\beta\partial
\varphi_{l'}/\partial t] \label{phase}\end{eqnarray}

with matrix $A$
\begin{eqnarray}
A=
\begin{pmatrix}
1+\alpha G&-\alpha&0&...&&&\\
-\alpha&1+2\alpha&-\alpha&0&...&&\\
0&-\alpha&1+2\alpha&-\alpha&0&...&\\
...&...&...&...&...&...&...\\
&&&...&0&-\alpha&1+\alpha G
\end{pmatrix}
\label{A-matrix}
\end{eqnarray}
where $l'$ runs over all  $N$ junctions, parameter $\alpha$  gives the coupling between junctions, $\beta$ is
dissipation parameter ($\beta^2 = 1/\beta_c$, where $\beta_c = \omega_p ^2 R^2 C^2$ is McCumber parameter,
$\omega_p$ is the plasma frequency and  $C$ is the capacity of the junction), $I$ is external current normalized
to the critical current $I_c$, $G=1+\gamma$, $\gamma=s/s_0=s/s_N$ and $s$, $s_0$, $s_N$ are the thickness of the
middle, first and last $S$-layers, respectively. In the equation ~(\ref{phase}) time is normalized to the plasma
frequency $\omega_p$.\cite{matsumoto99} According to the proximity effect we consider that the thickness of the
first and last layers are different from the layers inside of the stack. Nonperiodic boundary conditions (BC)
are characterized by parameter $\gamma$ and the equations for the first and last layers in the system
~(\ref{phase}) are different from the equation for the middle S-layer.\cite{koyama96,matsumoto99} For periodic
BC the matrix A has the form
\begin{eqnarray}
A=
\begin{pmatrix}
1+2\alpha&-\alpha&0&...&&&-\alpha\\
-\alpha&1+2\alpha&-\alpha&0&...&&\\
0&-\alpha&1+2\alpha&-\alpha&0&...&\\
...&...&...&...&...&...&...\\
-\alpha&&&...&0&-\alpha&1+2\alpha
\end{pmatrix}
\label{A-matrix-p}
\end{eqnarray}

We solve this system for the stacks with different number $N$ of intrinsic junctions. The numerical procedure
has been done as follows. For a given set of model parameters $N, \alpha, \beta, \gamma$ we simulate the IVC of
the system, i.e. $V_l(I)$, increasing $I$ from zero up and then down. A change in the parameters $N, \alpha,
\beta, \gamma$ change the branch structure in the IVC essentially. Their influence on the IVC in the CCJJ and
CCJJ+DC models was discussed in Refs.\onlinecite{{sm-physC1},{sm-physC2},{matsumoto99}}. To calculate the
voltages $V_l(I)$ in each point of IVC (for each value of $I$), we simulate the dynamics of the phases
$\varphi_l(t)$ by solving system of equations ~(\ref{phase}) using fourth order Runge-Kutta method. After
simulation of the phase dynamics we calculate the dc voltages on each junction as

\begin{eqnarray}
\partial\varphi_l/\partial t =\sum\sb{l'}\,A_{ll'}V_{l'} \label{volt}\end{eqnarray}
where $V_l$ is normalized to the $V_0=\hbar\omega_p/(2e)$. The average of voltage $\bar{V}_l$ is given by
\begin{eqnarray}
\bar{V}_l=\frac{1}{T_{max}-T_{min}}\int^{T_{max}}_{T_{min}}V_ldt
 \label{av-volt}\end{eqnarray}

where $T_{min}$ and $T_{max}$ determine the interval for the averaging. After completing the voltage averaging
for current $I$ , the current $I$ is increased or decreased by a small amount $\delta I$ to calculate the
voltages in the next point of the IVC. We use a distribution of phases and their derivatives achieved in the
previous point of the IVC as an initial distribution for the current point.

Numerical stability was checked by doubling and dividing in half the temporal discretization steps Dt and
checking the influence on the IVC.  Finally we can obtain the total dc voltage $V$ of the stack by

\begin{eqnarray}
V=\sum^{N}_{l=1}\bar{V}_l \label{tot-volt}
\end{eqnarray}

At some current $I$ some junction (or junctions) switches to the nonzero voltage state and gives some branch of
the IVC. We plot the total IVC at different parameters of the problem. The details concerning the numerical
procedure are given in Refs.\onlinecite{machida99,matsumoto99}.

To investigate the BPR in detail, we have calculated the IVC for different boundary conditions for the stacks
with different number N of IJJ from $N = 3$ to $N =30$. For clarity we restrict the number of curves in some
figures.

\section{ IVC for the stacks with different number of IJJ.}

\begin{figure}[!ht]
 \centering
\includegraphics[height=60mm]{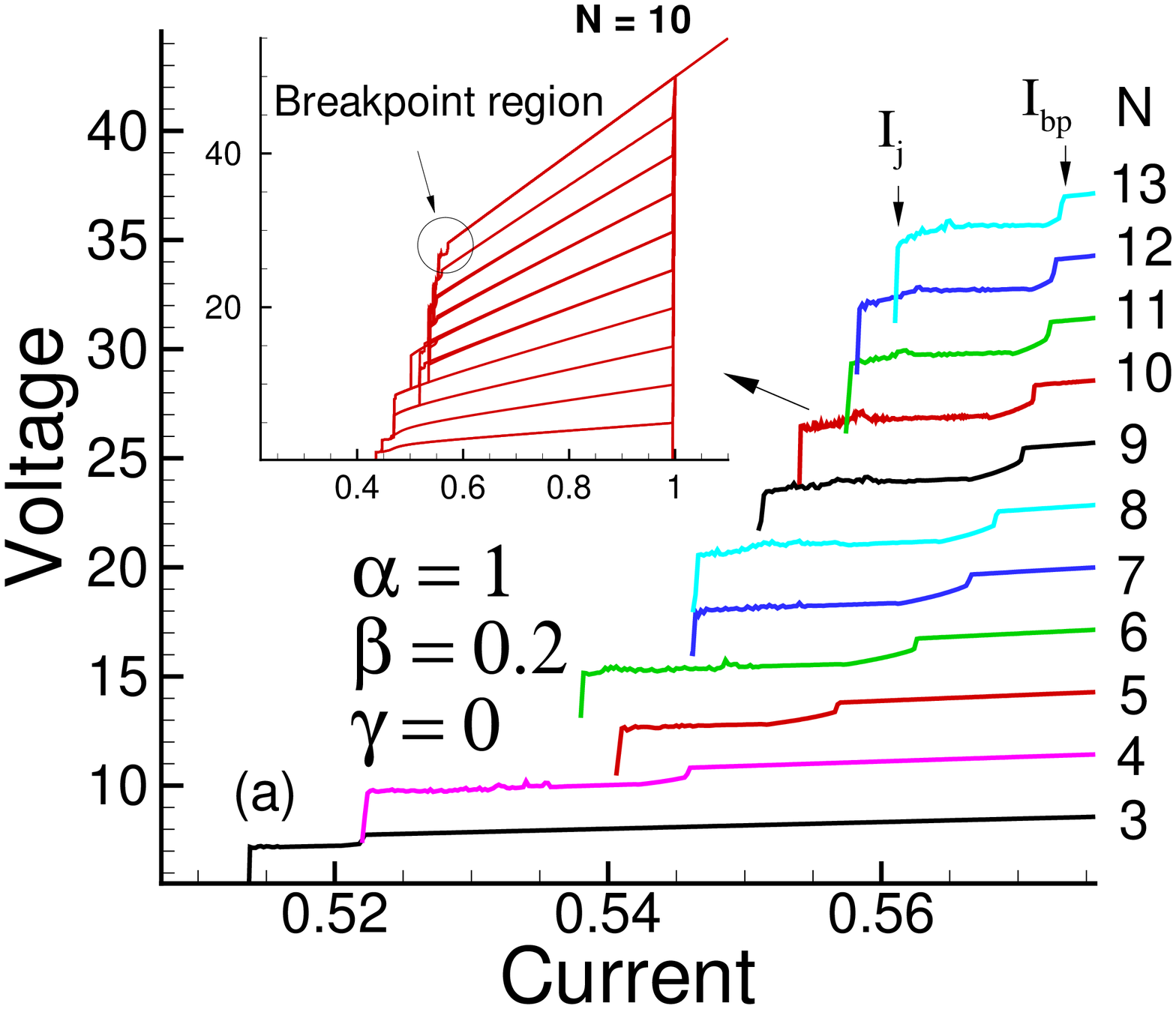}
\includegraphics[height=60mm]{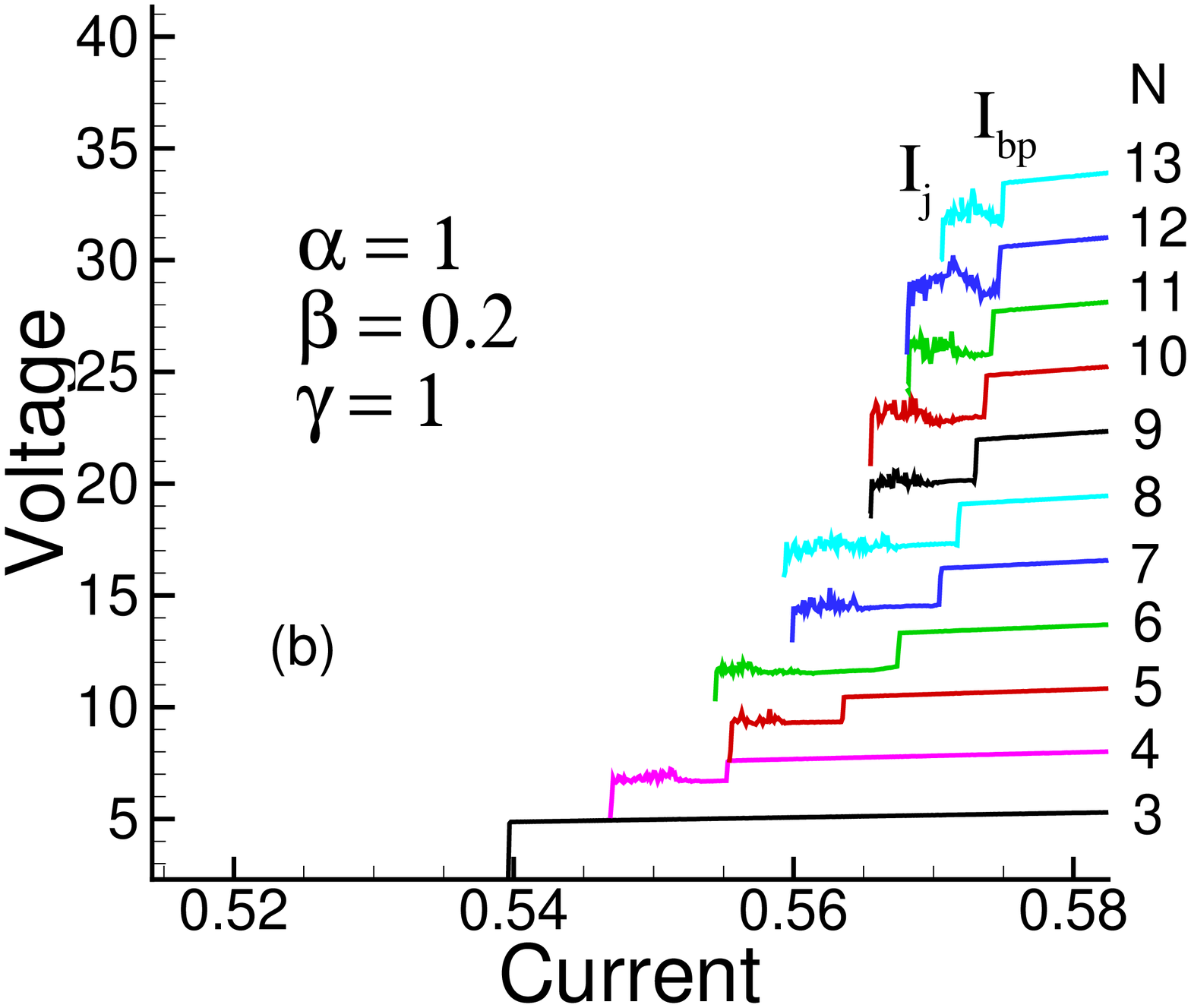}
\caption{(Color online) (a) - IVC of the outermost branch for the stacks with different number N of IJJ  at
$\gamma=0$; (b) - the same at $\gamma = 1$.} \label{1}
\end{figure}

Result of simulation of the total branch structure in the IVC for the stack of 10 IJJ in the CCJJ+DC model by
the equation (\ref{phase})  is presented in the insert of Fig.~\ref{1}a. As we can see, this IVC demonstrates
the BPR on the outermost branch.

 The outermost branch corresponds to the state of the stack with all junctions in the rotating
state (R-state)\cite{matsumoto99} and it is the upper branch in the IVC. The values of the breakpoint current
$I_{bp}$ and transition current $I_j$ (the jump point to the next branch in the IVC) on the outermost branch are
shown by arrows in Fig.~\ref{1}a. The distance between these two values we call as the width $w_{bp}$ of the
BPR.  We have found that the breakpoint current $I_{bp}$ and BPR width $w_{bp}$  depend on the parameters
$\alpha$ and $\beta$, boundary conditions and number of junctions in the stack.

Let us first describe the main features of the BPR which follow from the results of the simulation. As we can
see in Fig.~\ref{1}a, at  $\gamma = 0$ both $I_{bp}$ and $I_j$ are increased with N, but the increase of the
$I_{bp}$ is monotonic.   The IVC of the stacks with even N have larger $w_{bp}$ at small $N$. The IVC in the BPR
shows a chaotic behavior and its width is  decreased with $N$. There is a saturation of N-dependence of the
$I_{bp}$ at large N.

At  $\gamma = 1$ (Fig.~\ref{1}b) these features remain unchanged but the value of the $w_{bp}$ is decreased for
all stacks, especially for $N = 4$ and it is equal to zero for $N = 3$. This change in the boundary conditions
leads to the relative changing of the BPR width in different stacks as well.

\begin{figure}[!ht]
 \centering
\includegraphics[height=60mm]{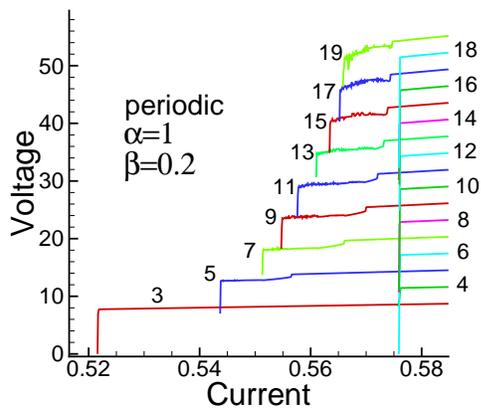}
\caption{(Color online) IVC of the outermost branch for the stacks with different number of junctions  at
periodic BC.} \label{2}
\end{figure}

The IVC at periodic boundary condition (Fig.~\ref{2}) show the same behavior for the $I_{bp}$ and BPR width
$w_{bp}$ for the stacks with odd N as in the nonperiodic case, but for the stacks with even N the value of the
$I_{bp}$ does not depend on the N and the BPR for these stacks is absent.

\begin{figure}[!ht]
 \centering
\includegraphics[height=60mm]{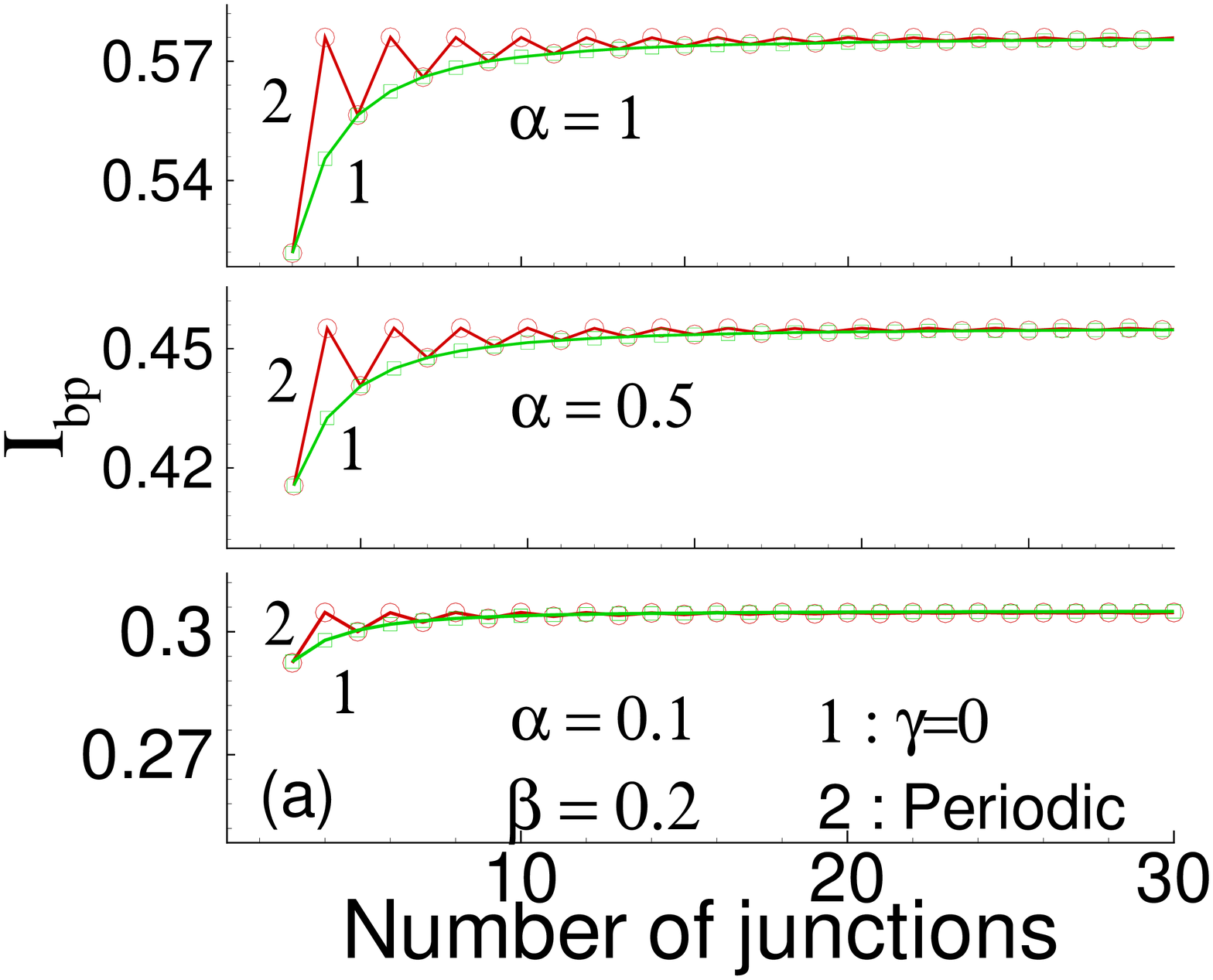}
\includegraphics[height=60mm]{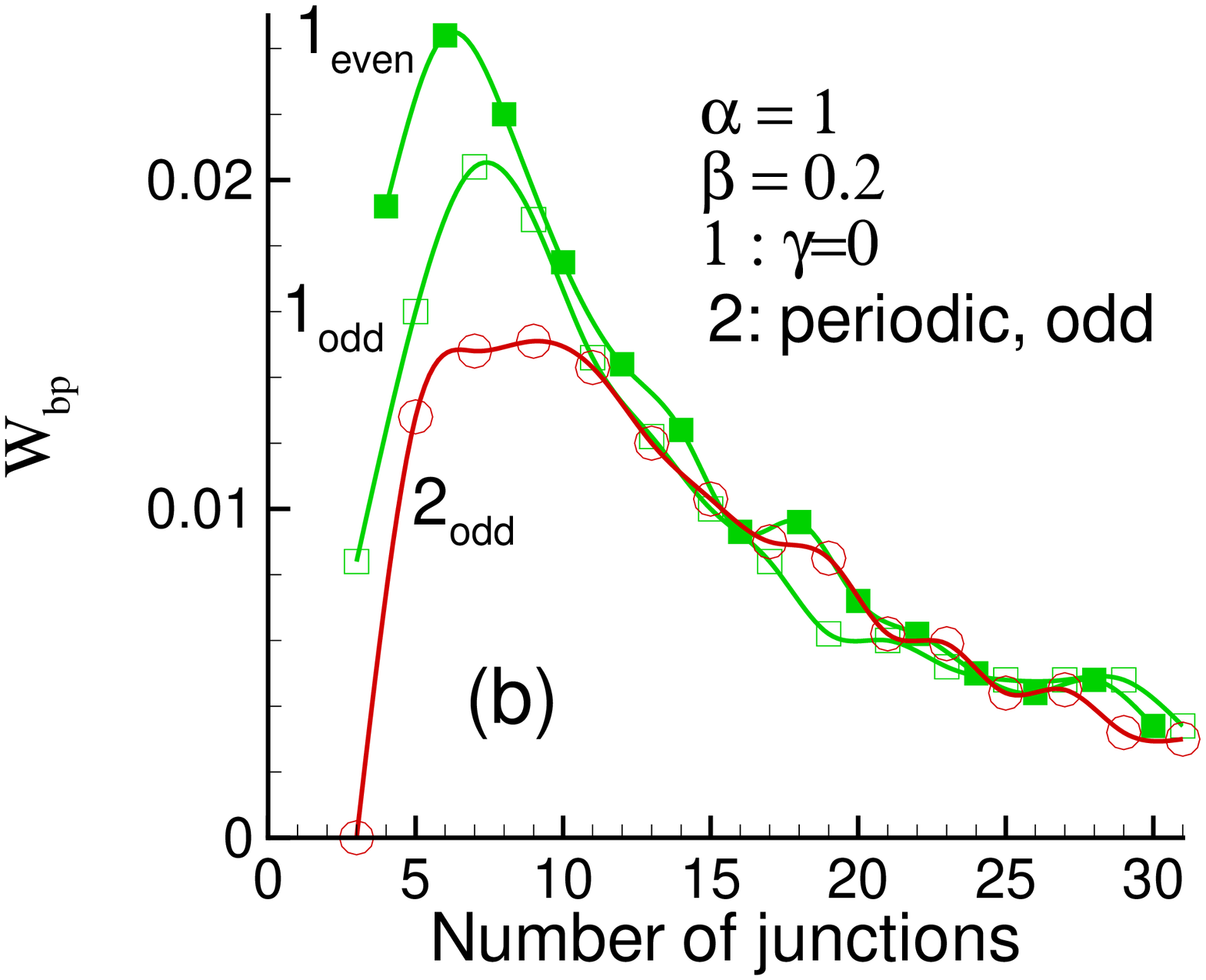}
\caption{(Color online) (a) -   the N-dependence of the $I_{bp}$ for $\gamma = 0$ (curve 1) and periodic (curve
2) boundary conditions at different $\alpha$; (b) -  the N-dependence of the BPR width $w_{bp}$ for stacks with
even and odd number of junctions at $\gamma = 0$ (1) and periodic (2) boundary conditions.} \label{3}
\end{figure}

In Fig.~\ref{3}a the $I_{bp}$ as a function of $N$ for $\gamma = 0$ (squares, curve 1)
and periodic (circles, curve 2) boundary conditions  at different value of the coupling
parameter $\alpha$ is presented. We stress the coincidence of the N-dependencies of the
$I_{bp}$ for stacks with odd numbers of IJJ for periodic and $\gamma=0$ cases. The
increase in $\alpha$ leads to the saturation of the N-dependence at larger N. The value
of saturated $I_{bp}$ is decreased with coupling and consists of 0.576 at $\alpha=1$,
0.454 at $\alpha=0.5$ and 0.304 at $\alpha=0.1$. At $\alpha = 0$ the breakpoint
coincides  with the return current, so the $I_{bp}$ has the same value for the stacks
with different number of junctions.

The N-dependence of the BPR width $w_{bp}$ for stacks with even and odd number of junctions  at $\gamma = 0$
(curves $1_{even}$ and $1_{odd}$) and periodic (curve $2_{odd}$) boundary conditions is shown in Fig.~\ref{3}b.
The main feature here is a decrease of the BPR width $w_{bp}$ with N at large N. At small N in the interval
(3,6) we observe the increase of the $w_{bp}$ with N.

\section{The origin of the breakpoint on the outermost branch}

To explain the observed features of the finite stacks IVC  let us discuss the origin of the breakpoint on the
outermost branch.  The hysteresis jump in the IVC is associated with the change of the distribution pattern of
rotating phase motions.\cite{matsumoto99} But  the question "why does a change in the current leads to the
change in a distribution pattern of the rotating phase motions" is still open. We consider a case that all
junctions are in the rotating state, i.e. the time average of $\bar{\varphi}_l$
($\bar{\varphi}_l=\frac{1}{T_{max}-T_{min}}\int^{T_{max}}_{T_{min}}\varphi_ldt$) is constant and that of $\sin
\varphi_l$ is zero for these junctions. For the oscillating junctions the situation is opposite: the time
average of $\bar{\varphi}_l$ is zero and that of $\sin \varphi_l$ is constant.

As we mentioned above, the outermost branch in the IVC corresponds to the state of the stack with all junctions
in the rotating state. Let us write an equation  for the difference of phase differences
$\delta_l=\varphi_{l+1}-\varphi_{l}$ for the outermost branch.

By subtracting equation ~(\ref{phase}) for $(l+1)th$ and $(l)th$ junctions we get
\begin{eqnarray}
(\ddot{\varphi}_{l+1}-\ddot{\varphi}_{l})+(1-\alpha\nabla^{(2)})\{\sin(\varphi_{l+1})-
\sin(\varphi_{l})\nonumber \\+\beta(\dot{\varphi}_{l+1}-\dot{\varphi}_{l})\}=0 \label{sub-d-e}
\end{eqnarray}
Here $\nabla^{(2)}f_l=f_{l+1}+f_{l-1}-2f_{l}$ is the discrete Laplacian. Consider a linear approximation
$\sin(\varphi_{l+1})- \sin(\varphi_{l}) \approx  \delta_l \cos(\varphi) $, where $\varphi\simeq\Omega t =
\frac{1}{N}Vt$,  $\Omega$ is Josephson frequency, $V$ is total voltage of the stack, we obtain
\begin{equation}
\ddot{\delta}_l+(1-\alpha\nabla^{(2)})(\cos(\varphi)\delta_l +\beta\dot{\delta}_l)=0\label{d-e-dp-l}
\end{equation}

\begin{figure}[!ht]
 \centering
\includegraphics[height=60mm]{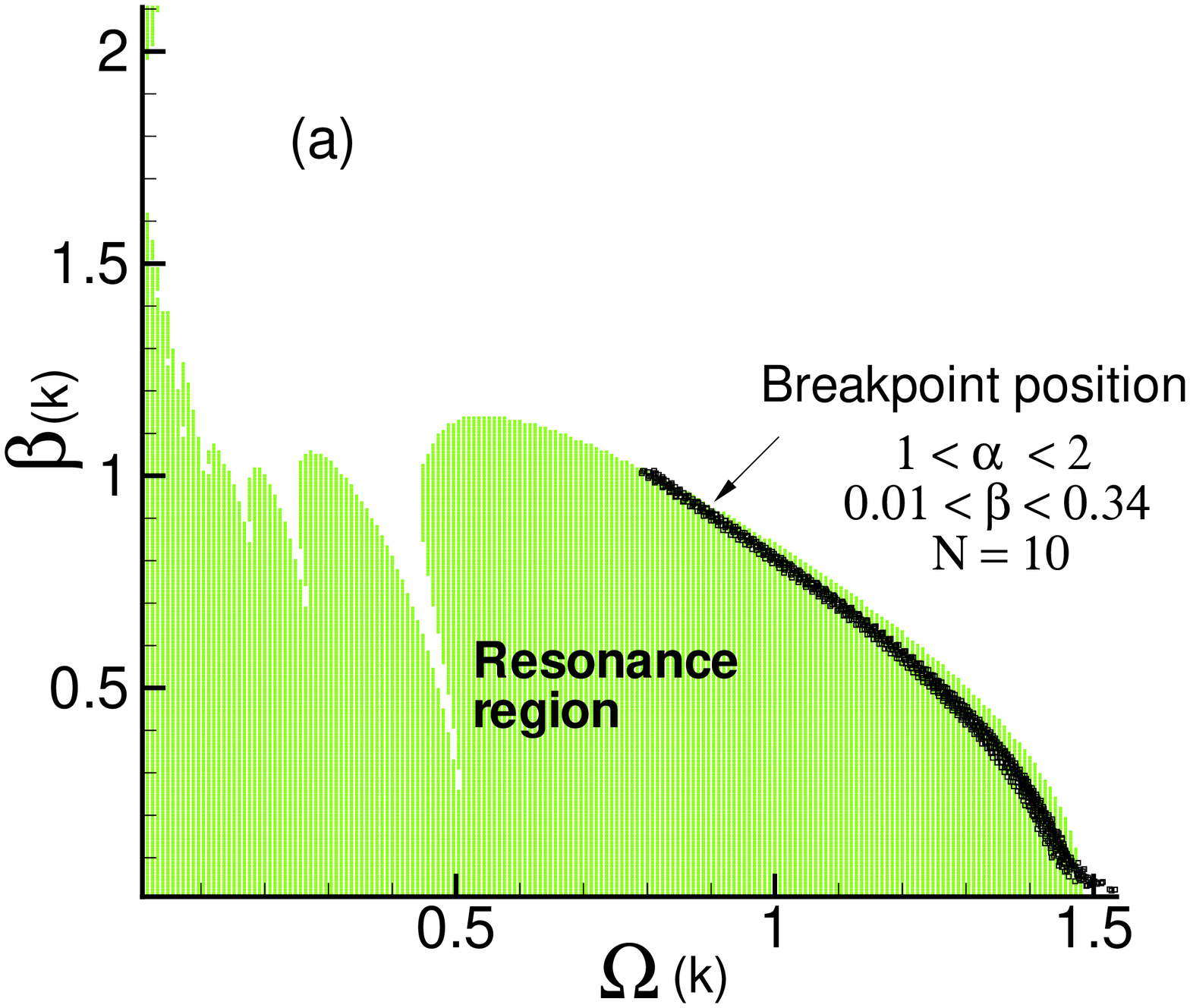}
\includegraphics[height=60mm]{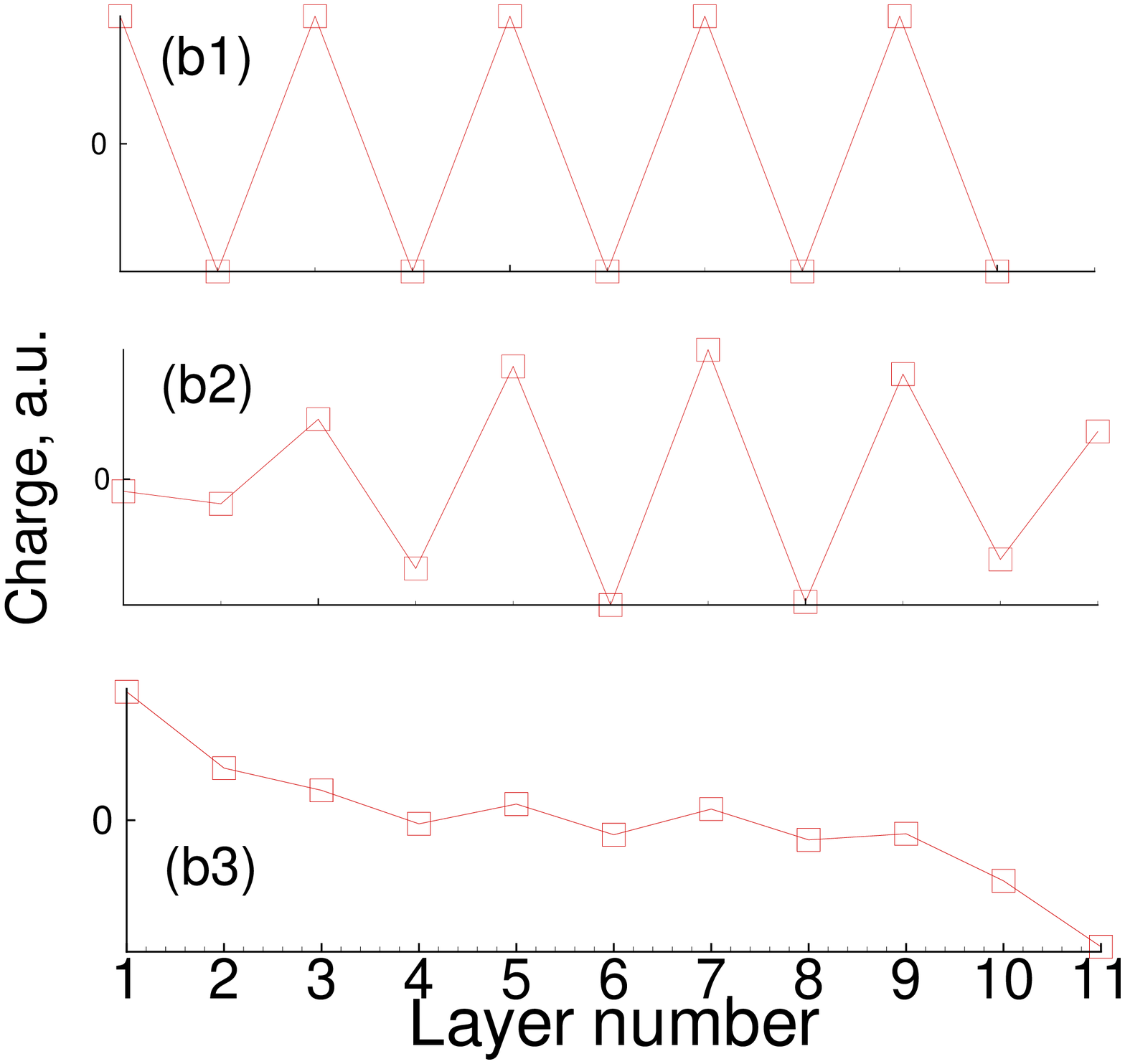}
\caption{(Color online) (a) - Parametric resonance region in $\Omega(k)-\beta(k)$ diagram. Black dots (stripe)
correspond to the  breakpoint current $I_{bp}$ in the IVC for $k=\pi$ at different values of parameters $\alpha$
and $\beta$; (b) - Charge on the S-layers  in the stacks with 10 IJJ (b1) and  11 IJJ (b2) at periodic BC and
with 10 IJJ at $\gamma = 1$ (b3). } \label{4}
\end{figure}

Expanding $\delta_l(t)$ in the Fourier series
\begin{equation}
\delta_l(t)=\sum_k\delta_k e^{ikl}
\end{equation}
the linearized equation for the Fourier component of a difference of the phase differences $\delta_k $ between
neighbor junctions can be written in the form\cite{sm-sust1}
\begin{eqnarray}
\ddot{\delta_k}+\beta(k)\dot{\delta_k}+ \cos(\Omega(k)\tau)\delta_k=0, \label{fur}
\end{eqnarray}
where $\tau=\omega_p(k)t$, $\omega_p(k)=\omega_p C_{\alpha}$, $\beta(k)=\beta C_{\alpha}$,
$\Omega(k)=\Omega/C_{\alpha}$ and $C_{\alpha} = \sqrt{1+2\alpha(1-\cos(k))} $.

The important fact for us is that this  linearized equation manifests a parametric resonance in the system of
IJJ. In Fig.~\ref{4}a we plot the resonance region for this equation on the diagram $\beta(k)-\Omega(k)$. The
dark stripe on this figure is actually the distribution of the dots, corresponding to the positions of the
breakpoints of the outermost branch.  Using the breakpoint values of the voltage in the equation
$\Omega(k)=\Omega/C_{\alpha}= V/NC_{\alpha}$ we obtain this distribution of the breakpoints  by the variation of
the coupling parameter $\alpha$ in the interval (1,2) with a step 0.1 and the dissipation parameter $\beta$ in
the interval (0.01,0.34) with the step 0.01 at each value of $\alpha$. In contrast to the results presented in
Ref.\onlinecite{{sm-sust1}} where the positions of the dots on the diagram $\beta(k)-\Omega(k)$ were obtained by
crude estimation, here we have done the precise numerical calculations. This calculations show more close
positions of the breakpoints to the boundary of the resonance region in the chosen intervals of $\alpha$ and
$\beta$.  The reason, that the position of the breakpoints are shifted from the boundary of the resonance region
in Fig.~\ref{4}a, is the linear approximation used to obtain the equation ~(\ref{d-e-dp-l}).

The breakpoints are inside of the resonance region, i.e. the resonance between the Josephson and plasma
oscillations is approached at the breakpoint $I_{bp}$.  As a result the plasma mode is excited by the Josephson
oscillations. We can prove this statement directly. By Maxwell equation, $div (E/d)=4\pi \rho$, we express the
charge $\rho_l$ on the S-layer $l$ by the voltages $V_{l,l-1}$ and $V_{l,l+1}$ in the neighbor insulating layers
$\rho_l=\frac{\epsilon_0}{4\pi d_0d}(V_{l, l+1}-V_{l-1 ,l})$. The time dependence of the $\rho_l$, presented in
Fig.~\ref{4}b1, demonstrates that at periodic BC in the stacks with 10 IJJ  the LPW with  $k =\pi$ is realized.
Really, the charge on the nearest neighbor layers has the same value and an opposite sign. Fig.~\ref{4}b2 shows
the distribution of the charge on the S-layers in the stacks with  11 IJJ  at periodic BC. In this case we
observe the creation of the LPW with $k =10\pi/11$. To determine the mode of the LPW at nonperiodic BC need more
detail investigation. Fig.~\ref{4}b3 demonstrates the charge distribution in the stack with 10 IJJ at $\gamma =
1$ near the breakpoint.

The wavelengths of the standing LPW which can be realized in the stack with $N$ junctions are $N/n$ units of
lattice in z-direction, where $n$ changes from 1 to $N/2$ for stacks with even number of junctions and from 1 to
$(N-1)/2$ for odd $N$. The voltage of the stack at the breakpoint is related to the wave number $k$ of the LPW
by formulae $V = N \Omega(k) \sqrt{1+2\alpha(1-\cos(k))} $, so the largest breakpoint voltage $V$ in the current
decreasing process corresponds to the creation of the LPW with $k$ equal to $\pi$ ($\pi$-mode) for stacks with
even number of IJJ and modes with $k=(N-1)\pi/N$ for stacks with odd N.

\section{Discussion of the main results }

Let us now return to the results presented in Fig.~\ref{1}-Fig.~\ref{3} and demonstrate that they are in
agreement with the ideas stated in the previous section.

According to these ideas, in the stack with 10 IJJ the LPW with $k=(N-1)\pi/N$ is created and it leads to the
increase of the  $I_{bp}$ with $N$ and its saturation to the value, corresponding to the $\pi$-mode. The same
modes are created in the stacks with $\gamma=1$ which outermost branches of the IVC are presented in
Fig.~\ref{1}b. They demonstrate the saturation of the $I_{bp}$ to the same value.

As Fig.~\ref{2} shows,  at periodic BC we observe the same value of $I_{bp}$ in the all stacks with even N. It
is in agreement with our suggestion  that  in this case the LPW with $k = \pi$ is created. We check it directly
as well, by the time dependence of the $\rho_l$. We find that at periodic BC in the stacks with even N the
charge on the nearest neighbor layers has the same value and an opposite sign which means that the LPW with $k
=\pi$ is realized.

In the stacks with odd N the $\pi$-mode cannot exist, so as we mentioned in the previous section, the LPW with
the largest $k$ equal to $k=(N-1)\pi/N$ is created. The creation of different modes of the LPW leads to the
different $I_{bp}$ and this fact explains the difference in the IVC at periodic BC of the stacks with even and
odd numbers of IJJ. With increase in N the wave number $k$ limits to $\pi$ and it leads to the increase in
$I_{bp}$ which we observe in Fig.~\ref{2}.

It explains as well the saturation of the $I_{bp}$ to the value corresponding to the $I_{bp}$ for stacks with
even N which is demonstrated in Fig.~\ref{3}a.

The difference between the charge distribution on the S-layers at the breakpoint current at fixed time for
stacks with even and odd numbers of IJJ (10 and 11)  at periodic BC is demonstrated in Fig.~\ref{4}b. In the
stack with odd N the charge on the first and last layers oscillates in-phase and the oscillations on the
neighbor S-layers are different from the oscillations in $\pi$-mode. Because at  $\gamma = 0$ the charge on the
first and last layers is screened due to the proximity effect, the LPW with the $k=(N-1)\pi/N$ as at periodic BC
for stacks with odd N is also  created both for stacks with odd and even N. This is a reason the values of
$I_{bp}$ for stacks with odd N coincide for both these BC as shown in Fig.~\ref{3}a.

At $\gamma =1$ the same modes are created, but the character of the charge distribution among the S-layers is
different (as we can see in Fig.~\ref{4}b3 ). As a result in this case the N-dependence of the $I_{bp}$ is
stronger than at $\gamma = 0$, but at $N\rightarrow \infty$ it is saturated as well and the saturation value is
the $I_{bp}$ for even junctions at periodic BC.

The influence of the coupling parameter on the value of the breakpoint current $I_{bp}$ and its N-dependence
which is demonstrated in Fig.~\ref{3}a has a clear explanation. It consists in the mentioned in the previous
section the $\alpha$-dependence of the voltage, and correspondingly, the breakpoint current $I_{bp}$. They are
proportional to the term $\sqrt{1+2\alpha(1-\cos(k))}$. An increase in $k$ leads to the term proportional to the
 $\sqrt{1+4\alpha}$. The decrease in $\alpha$ makes this influence weaker.

We may explain an increase in $w_{bp}$ at small $N$, which is shown in Fig.~\ref{3}b by a commensurability
effect on the width of the BPR. At periodic BC $w_{bp} = 0$, if the wavelength $\lambda $ of the LPW is $\lambda
= n $, where $n = 2,3,4...$  lattice units in z-direction. As we can see in Fig.~\ref{2}, at $\lambda = 3$ (for
stacks with $N = 3$) and $\lambda = 2$ ($\pi$-mode), the IVC do not show the BPR. The creation of the LPW with
wavelength in the middle of the interval (2,3) should correspond to some maximum of the BPR width $w_{bp}$. In
the stack with 5 IJJ the LPW with $\lambda = 2.5$ ($k = 4\pi/5$) is created. It does not coincide exactly with
the result obtained by the simulation,  because we use for the explanation the linearized equation for
difference of phase differences and Fourier expansion in the finite stacks. The wave number $k$ is not well
defined in this case. With increase in N the wave number $k$ of LPW limits to $\pi$ and it explains the increase
in $I_{bp}$ and the decrease in $w_{bp}$ at large $N$ which demonstrated by Fig.~\ref{3}b. The change in the
boundary conditions changes the character of the charge oscillations on the S-layers. Particularly, at $\gamma =
1$ we observe a decrease in the $w_{bp}$ for all stacks. Our analysis shows that in stacks with even N at
nonperiodic BC the charge on the second and $N-1$ layers oscillate in-phase, but anti-phase in the stacks with
odd N. We consider that the value of the BPR width $w_{bp}$ depends on the character of the charge oscillations
on the S-layers.

Finally, we note that in the case of coupling between junctions the parameter $\beta$ cannot be determined in a
usual way  by the return current, because it depends now on two parameters, $\beta$ and $\alpha$. The dependence
of the $I_{bp}$ and BPR width $w_{bp}$ on the dissipation and coupling parameters opens an opportunity to
develop the new method for determination of these parameters for the stacks of IJJ. This question will be
discussed in detail somewhere else.

\section{One oscillating junction}

Let us now discuss briefly  the breakpoints on the other branches of the IVC.

\begin{figure} [!ht]
 \centering
\includegraphics[height=60mm]{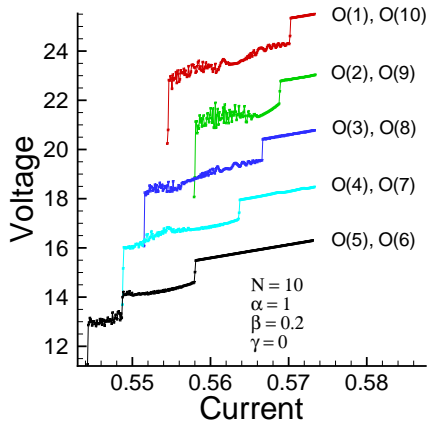}
\caption{(Color online) The BPR on the branches of the IVC of the stacks with one oscillating junction in the
case of 10 IJJ at $\alpha=1$, $\beta=0.2$ and $\gamma=0$. The upper curve corresponds to the real scale of
voltage, but the other ones are shifted for clarity by two units down.} \label{5}
\end{figure}

As we mentioned above a resistive state in the system of IJJ is realized as a state with different number of R-
and O-junctions.\cite{matsumoto99,sm-physC1} The different positions of R- and O-junctions in the stack
(different configurations) correspond to the different states of IJJ system.  The equidistant positions of the
O-junction from the ends of the stack (for example,  the stacks with 1st or 10th O-junction) lead to the same
state. So, there are five different states in the stack with one O-junction corresponding to the different
position of this junction. Fig.~\ref{5} shows the BPR on the branches of the IVC of the stacks with one
O-junction in the case of 10 IJJ at $\alpha=1$, $\beta=0.2$ and $\gamma=0$.

The equidistant positions of the O-junction from the ends of the stack  lead to the same value of $I_{bp}$ and
the same width of BPR.  The shift of the O-junction from the end of the stack to its center decreases the
$I_{bp}$ of the corresponding state. So, we may establish a delay of the LPW creation in the current decreasing
process when the position of the O-junction is shifted to the center of the stack.

We consider that the origin of such behaviour is the next. This one oscillating junction separates the stack in
two parts with different number of R-junctions which are weakly coupled through it. With a decrease in current
the first LPW is created in the part with the largest $I_{bp}$ (with the largest number of junctions). The shift
of the $O$-junction and the decrease in the number of R-junctions in this part lead to the decrease of the
$I_{bp}$ as Fig.~\ref{1} demonstrates. The increase of the number of junctions in the second part might manifest
the second breakpoint which is related to the creation of LPW in this second part of the stack. Such situation
is observed for $N=10$ when O-junction occupies the 5th or 6th site in the stack. The width of the BPR in the
other branches of IVC depends essentially on the state of the stack.

For the other branches the increase in the number of the O-junctions in the stack decreases the number of
effective junctions for creation of the LPW and it leads to the decrease of the return current. This fact
explains why we can obtain a total branch structure in the hysteresis region, because in the other case we would
not be able to observe it in the simulation. The correspondence between the position of the O-junction in the
stack and the value of the $I_{bp}$ opens the possibility for junction diagnostics, i.e. by measuring the value
of the $I_{bp}$ we can answer the question which junction in the stack goes into R- or O-state. From the other
side, the monitoring of the transitions between branches is useful for understanding the phase dynamics in the
system of IJJ.

\section{Conclusions}
In conclusion, we stress that the BPR in the IVC "naturally" follows from the solution of the system of the
dynamical equations for the phase difference for the stack of IJJ. In breakpoint region the plasma mode is a
stationary solution of the system and this fact might be used in some applications, particularly, in high
frequency devices such as THz oscillators and mixers. The detailed study of the breakpoint current and
breakpoint region width gives a new opportunity for the investigation of the properties of IJJ and and develop
new methods for the determination of the parameters of IJJ and diagnostic of IJJ in the stacks.

\section{Acknowledgments}
We thank P. M\" uller, R. Kleiner, A. Ustinov, T. Koyama, M. Machida, A. Yurgens, Yu. Latyshev, A. Irie, M.
Sargolzaei, T. Boyadjiev, N. M. Plakida, Y. Sobouti, M. R. H. Khajehpour for stimulating discussions and support
of this work.

\end{document}